\documentclass{aa}  
\usepackage{graphicx}
\usepackage{txfonts}
\usepackage{natbib}

\begin{document}

\title{Gemini/Phoenix H-band analysis of the globular cluster AL~3 
\thanks{Observations collected at the Gemini Observatory, Proposals GS-2006A-C9 and GS-2008A-Q-23-5, and at the European Southern Observatory (ESO), Proposal 64L-0212(A).}}

\author{
B. Barbuy\inst{1}
\and
H. Ernandes\inst{1,2,3}
\and
S. O. Souza\inst{1}
\and
R. Razera\inst{1}
\and
T. Moura\inst{1}
\and
J. Mel\'endez\inst{1}
\and
A. P\'erez-Villegas\inst{1}
\and
M. Zoccali\inst{4,5}
\and
D. Minniti\inst{6,7}
\and
B. Dias\inst{8}
\and
S. Ortolani\inst{9,10}
\and
E. Bica\inst{11}
}
\offprints{B. Barbuy}
\institute{
Universidade de S\~ao Paulo, IAG, Rua do Mat\~ao 1226,
Cidade Universit\'aria, S\~ao Paulo 05508-900, Brazil
\and
UK Astronomy Technology Centre, Royal Observatory, Blackford Hill, Edinburgh, EH9 3HJ, UK
\and
IfA, University of Edinburgh, Royal Observatory, Blackford Hill, Edinburgh, EH9 3HJ, UK
 \and
Instituto de Astrof\'{i}sica, Pontificia Universidad Cat\'{o}lica de
Chile, Av. Vicu\~{n}a Mackenna 4860, Santiago , Chile 
\and
Millennium Institute of Astrophysics, Av. Vicu\~{n}a Mackenna 4860,
782-0436 Macul, Santiago, Chile 
\and
Departamento de Ciencias Fisicas, Facultad de Ciencias Exactas, Universidad Andres Bello, Av. Fernandez Concha 700, Las Condes, Santiago, Chile
\and
Vatican Observatory, V00120 Vatican City State, Italy
\and
Instituto de Alta Investigaci\'on, Sede Esmeralda, Universidad de Tarapac\'a, Av. Luis Emilio Recabarren 2477, Iquique, Chile
\and
Universit\`a di Padova, Dipartimento di Fisica e Astronomia, Vicolo
 dell'Osservatorio 2, I-35122 Padova, Italy
\and
INAF-Osservatorio Astronomico di Padova, Vicolo dell'Osservatorio 5,
I-35122 Padova, Italy
\and
Universidade Federal do Rio Grande do Sul, Departamento de Astronomia,
CP 15051, Porto Alegre 91501-970, Brazil
}
\date{Received; accepted }
 \abstract
     {The globular cluster AL~3 is old and located in the  inner bulge. Three individual stars were observed with the Phoenix spectrograph at the Gemini South telescope.
       The wavelength region contains prominent lines of CN, OH, and CO, allowing the derivation of C, N, and O abundances of cool stars.}
{We aim to derive C, N, O abundances of three stars in the bulge globular cluster AL~3, and
  additionally in stars of NGC~6558 and HP~1. The spectra of AL~3 allows us to
   derive the cluster's radial velocity.}
   {For AL~3, we applied a new code to analyse its colour-magnitude diagram. Synthetic spectra were computed and compared to  observed spectra for the three clusters. }
   {We present a detailed identification of lines in the spectral region
   centred at 15555 {\rm \AA}, covering the wavelength range
   15525-15590 {\rm \AA}.  C, N, and O abundances are tentatively derived for the sample
   stars.}
  {}
\titlerunning{Phoenix analysis of the globular cluster AL~3 }
\authorrunning{B. Barbuy et al.}
\maketitle
%
\section{Introduction}

The globular
clusters  AL~3, NGC~6558, and HP~1 share the characteristics of
having a metallicity of [Fe/H]$\sim$-1.0 and of being located in the Galactic bulge. They are old and could represent the earliest stellar populations in the Galaxy (Ortolani et al. 2006; Barbuy et al. 2018a; Kerber et al. 2019). 

The star cluster AL~3 was discovered by Andrews \& Lindsay (1967) and was also cataloged as
BH~261 by van den Bergh \& Hagen (1975), reported as a faint open cluster. 
It is reported in the ESO/Uppsala catalogue (Lauberts 1982) as ESO 456-SC78.
Ortolani et al. (2006) showed that the star cluster shows B, V, I
colour-magnitude diagrams (CMD) typical of a globular cluster.
It is centred at
 J2000 $\alpha$ = 18$^{\rm h}$14$^{\rm m}$06.6$^{\rm s}$,
$\delta$ = $-$28$^{\rm o}$$38'06''$, with Galactic coordinates 
 l = 3.36$^{\rm o}$, b = $-$5.27$^{\rm o}$,
and located at 6\fdg25 and 2 kpc from
the Galactic centre, hence in the inner bulge volume. The cluster
has a depleted red giant branch (RGB), similarly to low-mass Palomar clusters,
indicating it to have been stripped along its lifetime.
This cluster has not been further observed so far.

The NGC~6558 cluster is located in a window, identified by Blanco (1988),
with equatorial coordinates (J2000) 
$\alpha$ = 18$^{\rm h}$ 10$^{\rm m}$18.4$^{\rm s}$, 
$\delta$ = -31$^{\circ}$ 45' 49" and
 Galactic coordinates  l = 0.201$^{\circ}$, b = -6.025$^{\circ}$.
It was analysed in terms of CMD by 
Rich et al. (1998). Rossi et al. (2015) obtained a proper-motion-cleaned
CMD and presented a proper motion analysis, from which a study
of its orbits was given in P\'erez-Villegas et al. (2018, 2020).

The globular cluster Cl Haute-Provence 1 or HP~1, also
designated BH 229 and ESO 455-SC11, was discovered by
Dufay et al. (1954). It is located at 
J2000 $\alpha$ = 17$^{\rm h}$31$^{\rm m}$05.2$^{\rm s}$, 
$\delta$ = $-$29$^{\rm o}$$58'54''$, with Galactic coordinates
l = 357.42$^{\rm o}$, b = 2.12$^{\rm o}$. 

In the present work, we studied individual stars of these clusters in a limited region of the spectrum in the H-band
corresponding to the wavelength region of the Phoenix spectrograph
at the Gemini South telescope, 
centred at 15555 {\rm \AA}, and covering 15520-15590 {\rm \AA}, with
a high spectral resolution of R$\sim$75,000.
This region was chosen for containing prominent lines of CN, CO, and OH.

In Sect. 2, we examine the list of atomic and molecular lines in the region.
In Sect. 3, the observations are described. In Sect. 4,
the CMD is analysed. In Sect. 5,
the stellar parameters are given. In Sect. 6, the C, N, O abundances
are derived. Conclusions are drawn in Sect. 7.

\section{Spectroscopy in the H-band: Atomic and molecular lines}

The H-band will be intensely observed in the near future, given the new
instruments placing emphasis on the near-infrared region, such as the
James Webb Space Telescope (JWST), and new spectrographs on ground-based telescopes such as MOONS@VLT (presently CRIRES@VLT is available) and MOSAIC@ELT. The project APOGEE (Apache Point Observatory Galactic Evolution Experiment), with observations at a resolution of R$\sim$22,000 carried out at the 2.5-metre Sloan Foundation Telescope at the Apache Point Observatory in New Mexico (APOGEE-2N), and the 2.5-metre du Pont Telescope at Las Campanas Observatory in Chile (APOGEE-2S), as described in Majewski et al. (2017) has shown the power of the H-band.

Given the short wavelength range covering only 70 {\rm \AA} of the Phoenix spectrograph, we faced the challenge of identifying the lines in moderately metal-poor
stars of the Galactic bulge, for which only a few lines
are available. Because the available lamps did not include lines in this region, and experience proved that sky lines  yielded a better wavelength calibration, and given the short wavelength range,  it is not straightforward to identify the lines.

For this reason, we proceeded to a line identification, in the spectra of
the reference stars Arcturus and $\mu$ Leo, and created a
shortened version of a line list, containing only detectable lines.

Mel\'endez \& Barbuy (1999, hereafter MB99) worked on a list of atomic lines in the
J and H bands. The list of lines mostly corresponded to the detectable
lines. 
That previous line list needed to be largely completed. Upon checking the lines detectable in the wavelength range 15520-16000 
\rm \AA, 
this was done by verifying the line lists from APOGEE (Shetrone et al. 2015) and VALD (Piskunov et al. 1995, Ryabchikova et al. 2015).
We note that astrophysical oscillator log~gf strengths
were applied to the APOGEE line list, and these should be suitable for abundance derivation.
Through a line-by-line check of its detectability in the Arcturus spectrum,
we identified lines of  \ion{Mg}{I}, \ion{Si}{I},
\ion{Ca}{I}, \ion{Ti}{I}, \ion{Mn}{I}, \ion{and Ni}{I}, 
and we were not able to find detectable lines from the species \ion{C}{I},
\ion{O}{I}, \ion{Sc}{I}, \ion{V}{I}, \ion{Cr}{I}, \ion{Co}{I}, \ion{Cu}{I},
\ion{Y}{I}, \ion{and Y}{II}. The spectra computed including all lines
of all these elements are entirely equivalent to the one computed with the shortened line list, therefore to ensure practicality when identifying which lines really contribute
to a feature, we created  
 a table containing the detectable lines only. In this table, available upon request,
we report the oscillator strengths from MB99, APOGEE, VALD, and adopted values,
where in order of preference we adopted NIST and MB99. 

Molecular electronic transition lines of CN A$^{2}\Pi$-X$^{2}\Sigma$, 
and vibration-rotation
CO X$^{1}\Sigma^{+}$, OH X$^{2}\Pi$ 
lines were included in the synthetic spectra
calculations.
The line lists for CN were made available by
S. P. Davis, the CO line lists were adopted from
Goorvitch (1994), and the OH line list was made available by
S. P. Davis and A. Goldman (Goldman et al. 1998).
For more details on CN, CO, and OH molecular lines, we refer the reader to
Mel\'endez \& Barbuy (1999), Mel\'endez et al. (2001), and
Mel\'endez et al. (2002).
TiO $\phi$-system b$^{1}$$\Pi$-d$^{1}$$\Sigma$ lines are also present in the region. The line list by Jorgensen (1994) is included in the
calculations as described in 
Schiavon \& Barbuy (1999) and Barbuy et al. (2018a).
The adopted dissociation potential of OH is 4.392 eV, D$_0$ = 11.092 eV for CO and D$_0$ = 7.65 
eV for CN (Huber \& Herzberg 1979).
The PFANT code employed here for calculations of synthetic spectra is described in Barbuy et al.
(2018b), and it is available together with the atomic and molecular line lists.\footnote{http://trevisanj.github.io/PFANT}  

We identified the lines in the reference stars Arcturus and $\mu$ Leo.
For the reference star Arcturus, 
the spectrum atlas from Hinkle et al. (1995) is used, 
and for the metal-rich reference giant star $\mu$ Leo,
  APOGEE spectra are used, and their studies are be presented elsewhere.

 The adopted stellar parameters for Arcturus and $\mu$ Leo
 are from Mel\'endez et al. (2003) and  Lecureur et al. (2007),
 and they are reported in Sect. 5.

\begin{table}
\small
\caption[2]{\label{logbook}
Log book of observations. Proposals GS-2006A-C9 on 15-16/07/2006,
and GS-2008A-Q-23 on 10/04/2008 and 22/05/2008}
\begin{flushleft}
\begin{tabular}{lrrrrrrrrrrrrrrrrrrrrrrrrrrrrrrrrrrr}
\noalign{\smallskip}
\hline
\noalign{\smallskip}
\hbox{Object} & Date  & LST & Exp.  & Seeing & \\
 &   & h:m & (m) & (") & \\
\noalign{\vskip 0.2cm}
\noalign{\hrule\vskip 0.2cm}
\noalign{\vskip 0.2cm}
NGC~6558-11 & 15-07-2006 & 22:00 & 2x25 & 0.6 & \\
NGC~6558-64 & 16-07-2006 & 14:48 & 3x30 & 0.7 & \\
NGC~6558-11 & 16-07-2006 & 16:33 & 2x20 & 0.5 & \\
NGC~6558-42 & 16-07-2006 & 19:13 & 2x30 & 0.7 & \\
NGC~6558-97 & 16-07-2006 & 20:25 & 3x20 & 0.6 & \\
NGC~6558-73 & 16-07-2006 & 21:38 & 2x30 & 0.9 & \\
HP~1-4 & 15-07-2006 & 13:42 & 3x30 & 0.6 & \\
HP~1-6 & 15-07-2006 & 20:12 & 3x30 & 0.6 & \\
HP~1-5 & 16-07-2006 & 13:03 & 3x30 & 0.8 & \\
HP~1-3 & 16-07-2006 & 18:02 & 2x30 & 0.6 & \\
AL~3-3 & 10-04-2008 & 15:58 & 4x15 & --- & \\
AL~3-6 & 22-05-2008 & 15:09 & 4x15 & --- & \\
AL~3-7 & 22-05-2008 & 16:21 & 4x15 & --- & \\
\noalign{\smallskip} \hline \end{tabular}
\end{flushleft}
\end{table}

\section {Observations} 

The spectra of  red giant stars of the bulge globular clusters
NGC~6558, AL~3, and HP~1 were observed with the Phoenix spectrograph
installed on the 8m Gemini South telescope. 
The program was tri-national from
Brazil (PI: B. Barbuy),  Chile (PI: M. Zoccali),
and Australia (PI: J. Mel\'endez).

The final suitable spectra include  three stars of AL~3, two stars of NGC~6558, 
and one star of HP~1, centred at 1.555 $\mu$m in the H band. 
Another three stars in NGC~6558, and three in HP~1
were also observed; however, these showed a low Signal-to-Noise (S/N), due to clouds or high airmass.
 The log of observations is provided in Table \ref{logbook}.
 
\begin{figure}
\includegraphics[angle=0,scale=.50]{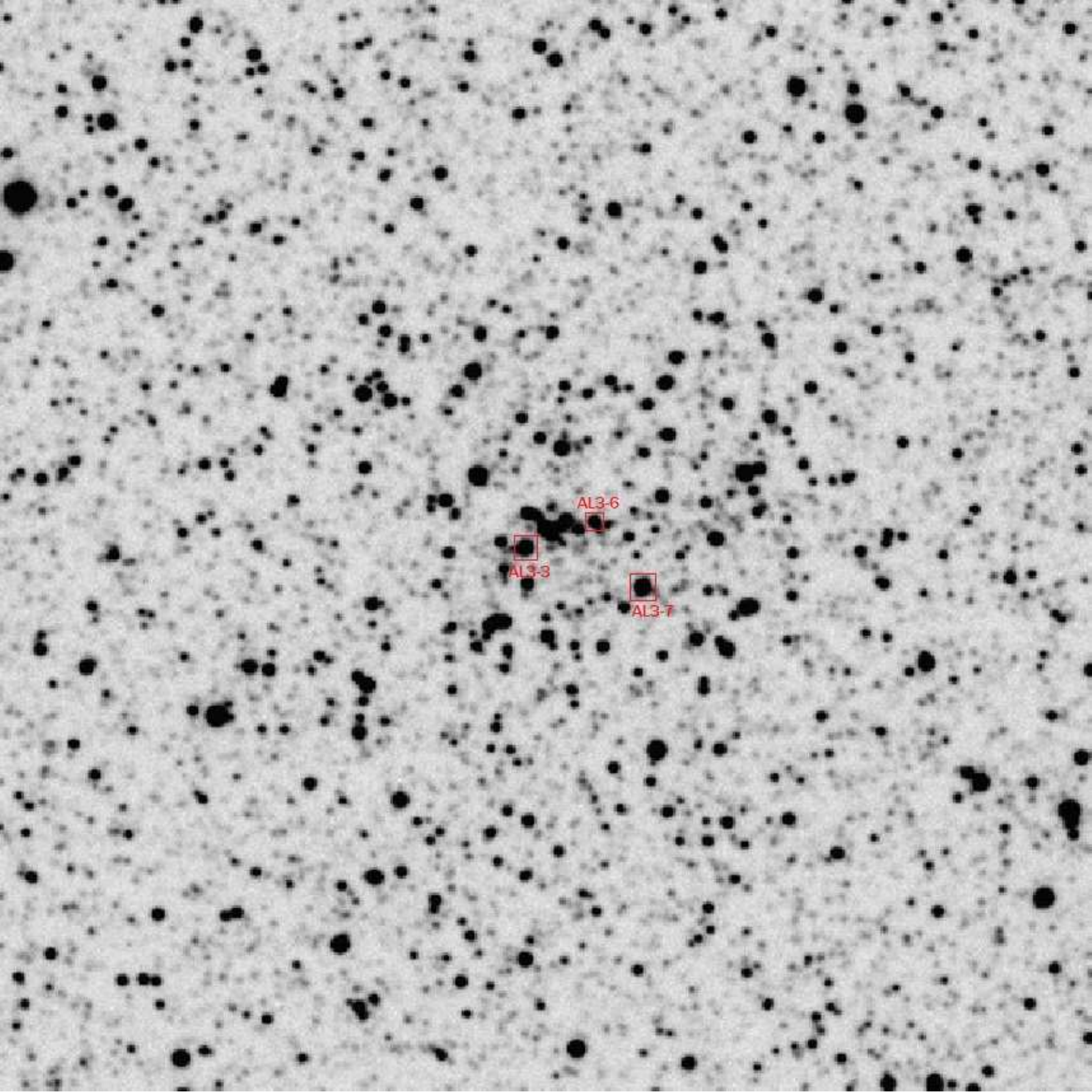}
\caption{AL~3: 3 min B image with the three sample stars identified.
 Extraction of 3'$\times$3'.
North is up and east is left.}
\label{chart}
\end{figure}

 This is the first spectroscopic work on AL~3, except for Gaia
 data (Gaia Collaboration 2020).
In Fig. \ref{chart}, a 3 min  $B$ exposure of AL~3
is shown for a field extraction of  3.3'$\times$3.3' 
(510$\times$510 pixels).  
The sample stars of AL~3 are identified in the chart.
Charts and identifications of the observed stars in NGC~6558 and
HP~1 are given in Barbuy et al. (2009, 2018) and Barbuy et al. (2006,
2016), respectively.

\subsection{Gaia cross-check}


In order to verify the corresponding membership probability of observed stars in AL~3, we performed the cross-match with Gaia Early Data Release 3 (EDR3; Gaia Collaboration 2020). We selected stars within 20' from the cluster centre and used the renormalised unit weight error (RUWE) $\le 2.4$ to ensure the kinematics precision and the minimum match separation.

Having the high-precision EDR3 proper motions ($\mu_\alpha^*$ = $\mu_\alpha \cos \delta$ and $\mu_\delta$), we obtain the mean proper motions for the cluster of $\mu_\alpha^* =  3.59\pm0.03$ mas yr$^{-1}$ and $\mu_\delta = -3.54\pm0.04$ mas yr$^{-1}$. These values are compatible with those given in Baumgardt et al. (2019). We also computed the Gaussian membership probability distribution of AL~3. We found that the stars AL3-6 and AL3-7 have membership probabilities of 100\%. Finally, the star AL3-3 has a relatively low membership probability of $\sim$ 60\%, but still, it could be considered a member. Therefore, all three observed stars are probable members of AL~3. Table \ref{gaia} provides the Gaia EDR3 cross-match and the membership probabilities.

\begin{table*}
\small
\caption[2]{\label{gaia}
Gaia magnitudes, proper motions and membership probability.}
\begin{center}  
\begin{tabular}{lccccccc}
\noalign{\smallskip}
\hline
\noalign{\smallskip}
Star   &      $\alpha$(J2000)  &  $\delta$(J2000)  &   $G$  &  $G_{RP}$ &   $\mu_{\alpha}^*$ &  $\mu_{\delta}$ &    Memb \\
  & deg & deg & mag & mag & mas yr$^{-1}$ & mas yr$^{-1}$ & \% \\
\noalign{\smallskip}
\hline
\noalign{\smallskip}
    AL3-3  &  273.5288067 & -28.6357960 & $14.186\pm0.003$ & $13.323\pm0.004$ & $3.78\pm0.09$ & $-4.04\pm0.07$ & 59 \\
    AL3-6  &  273.5247404 & -28.6346067 & $15.228\pm0.003$ & $14.303\pm0.011$ & $3.56\pm0.06$ & $-3.64\pm0.04$ & 100 \\
    AL3-7  &  273.5220767 & -28.6380356 & $13.924\pm0.003$ & $13.062\pm0.005$ & $3.58\pm0.03$ & $-3.48\pm0.03$ & 100 \\
\noalign{\smallskip} \hline \end{tabular}
\end{center}
\end{table*}

\subsection{Radial velocity of AL~3}
\begin{figure*}[h!]
\centering
\includegraphics[angle=0,scale=.50]{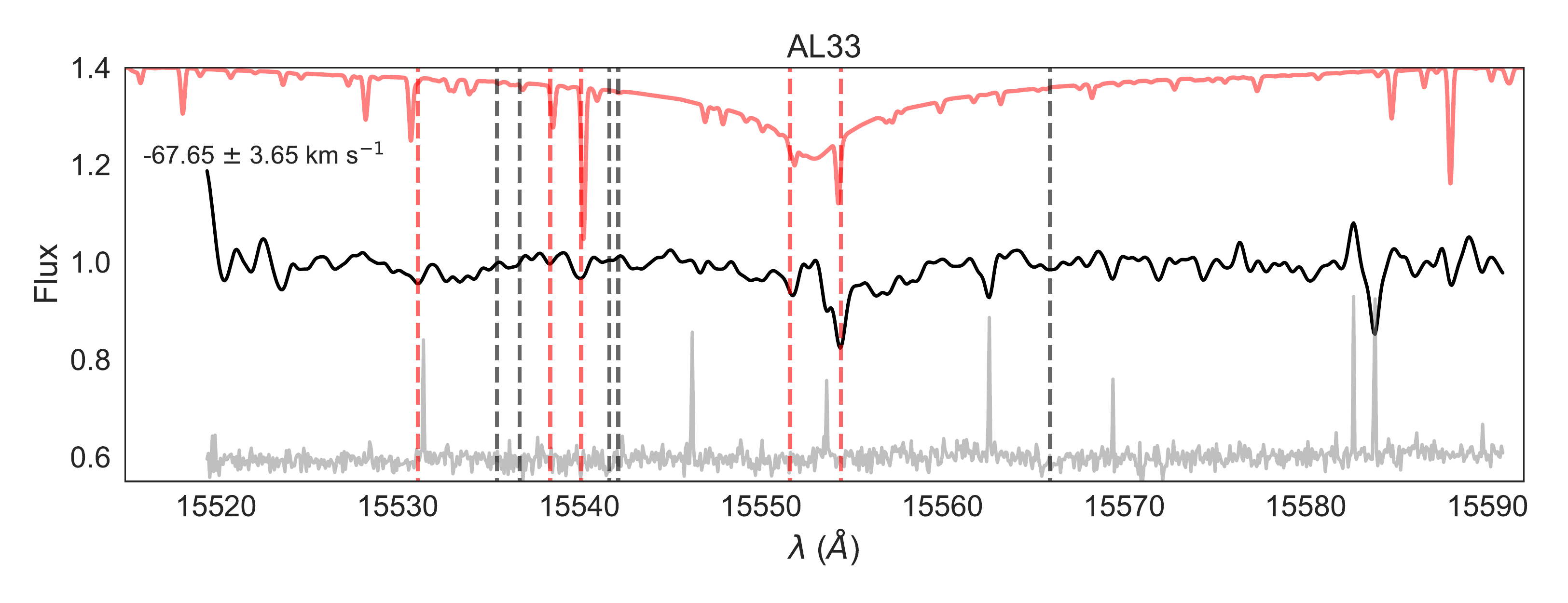}
\caption{AL~3-3: Radial velocity derivation. The solid black line is the observed spectrum, the solid grey line is the noise spectrum, the solid red line is the synthetic spectrum, the dashed red lines are those used to derive the radial velocity, and the dashed black lines are the OH sky lines.}
\label{rval33}
\end{figure*}

\begin{figure*}[h!]
\centering
\includegraphics[angle=0,scale=.50]{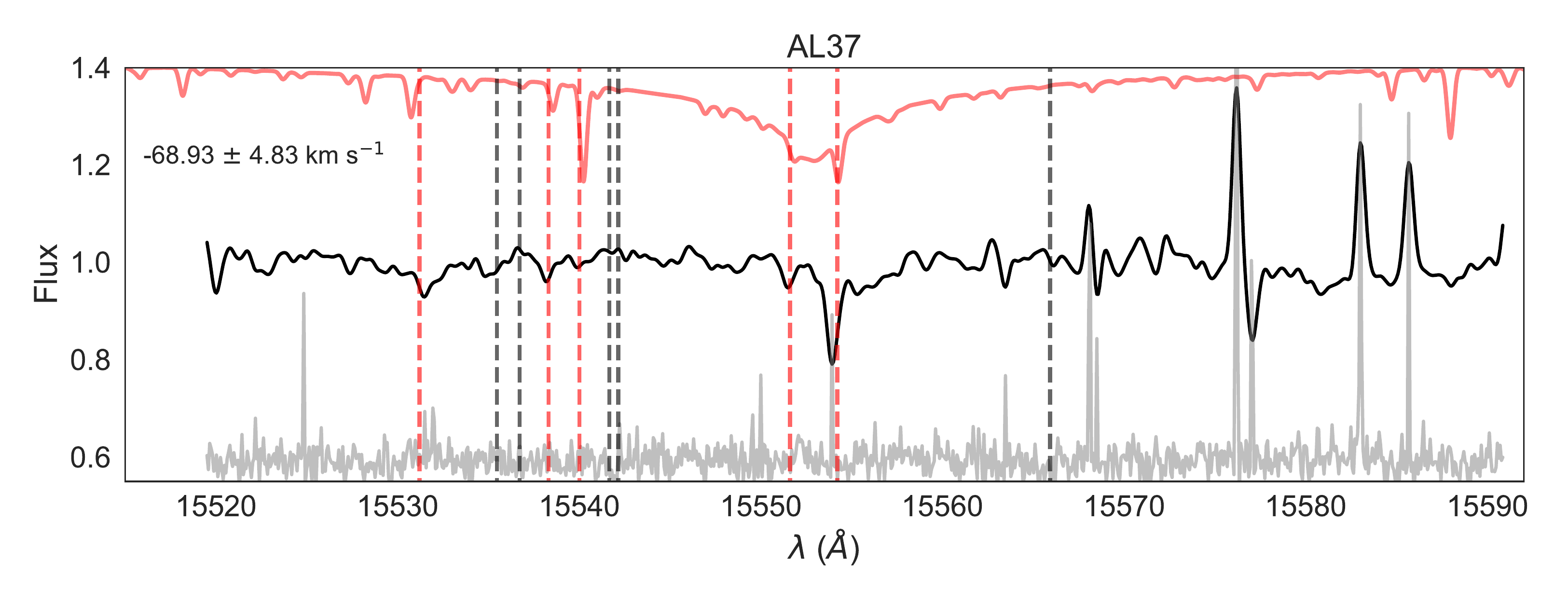}
\caption{AL~3-7: Radial velocity derivation. The solid black line is the observed spectrum, the solid grey line is the noise spectrum, the solid red line is the synthetic spectrum, the dashed red lines are those used to derive the radial velocity, and the dashed black lines are the OH sky lines.}
\label{rval37}
\end{figure*}

We were able to derive radial velocities
for the sample stars. 
We used the low S/N individual observations of each star (S/N$\sim$10.0) combined to increase the signal-to-noise to S/N$\sim$18.0, and S/N$\sim$22.0 for AL3-3 and AL3-7, respectively. 
Due to the stacking process, the most prominent features identified are FeI 15,534.26, OH 15,542.10, TiI 15,543.758, TiO/NiI 15,55.25 blend, CN 15,555.25, and FeI 15,591.49. We also used the OH sky lines, as listed
in Table 2 by Mel\'endez et al. (2003).
These features were used for AL3-3, giving a radial velocity of -67.65 $\pm$ 3.65 kms$^{-1}$. In the combined spectrum of AL3-7, the same features result in a radial velocity of -68.93 $\pm$ 4.83 kms$^{-1}$.  
The corresponding heliocentric velocities of -57.29 km.s$^{-1}$ and -58.57 km.s$^{-1}$ lead to a final mean heliocentric velocity of -57.93 km.s$^{-1}$  $\pm$ 4.28 for AL~3.
Figures \ref{rval33} and \ref{rval37} show the line identification and radial velocity derivation.

\begin{figure*}[h!]
\centering
\includegraphics[angle=0,scale=.50]{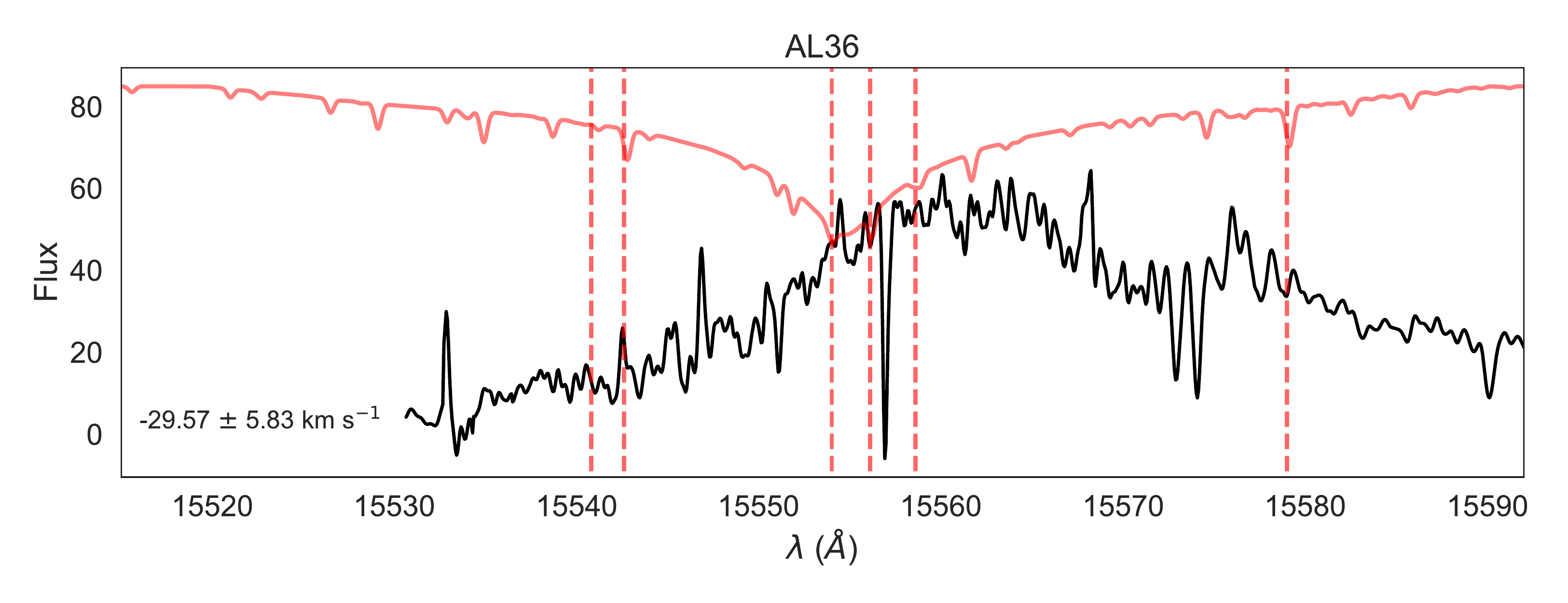}
\caption{AL~3-6: Radial velocity derivation. The solid black line is the observed spectrum, the
solid grey line is the spectrum of star AL3-7, the dashed red lines are those used to derive the radial velocity, and the dashed black lines are the OH sky lines.}
\label{rval36}
\end{figure*}

The star AL3-6 shows a very noisy spectrum, and we verified
that it was observed under a high airmass of over 1.3, which
also explains that it is plagued with telluric features.
 For AL3-6, we obtained  a 
different heliocentric radial velocity, as shown in Fig. \ref{rval36}, of -29.57 $\pm$ 5.85 km.s$^{-1}$,
compatible with the value given by Baumgardt et al. (2019) of -29.38 $\pm$ 0.60 km.s$^{-1}$.

The derived radial velocity is of crucial importance for
the computation of the cluster's orbits - see Sect. 5.
However, we 
obtained two different figures: v$_{\rm r}^{hel}$ = -57.93 and -29.57 
km.s$^{-1}$. In Fig. \ref{rval36}, we show the spectrum of AL3-6 compared with that of AL3-7.
Therefore, we obtained two different radial velocities for
AL~3, and consequently we considered both of them for the
calculation of orbits.
Since we had already computed the orbits for the lower value
(from Baumgardt et al. 2019), we show the orbits for the higher value here.
Clearly, new observations of these stars in a more extended wavelength range and with a higher S/N would be of great interest.

\section{Colour-magnitude diagram of AL~3}

 Ortolani et al. (2006) presented $B$, $V$, and Cousins $I$ images of AL~3, observed on 2000 March 6 using the 1.54m Danish telescope at the European Southern Observatory (ESO) at La Silla. They derived a reddening of E(B-V) = $0.36\pm0.03$ and a distance of $6.0\pm0.5$ kpc, based on the magnitude of the horizontal branch (HB) of AL~3, and suggested a metallicity of [Fe/H]$=-1.3\pm0.3$ from a comparison of the CMD with the mean locus of the cluster M5. 

Rossi et al. (2015) described the observations with EFOSC at the NTT (ESO), La Silla in May 2012. They reported a distance of $\sim 6.5$ kpc, in agreement with Harris (1996, 2010 edition). P\'erez-Villegas et al. (2020) adopted both distances of 6.0 and 6.5 kpc and the radial velocity from Baumgardt et al. (2019) for the calculation of orbits.

We carried out the isochrone fitting using the SIRIUS code (Souza et al. 2020) on the same AL~3 data presented in Ortolani et al. (2006). The metallicity was limited by using a Gaussian prior with the value of Ortolani et al. (2006). We adopted 
the Dartmouth Stellar Evolution Database (DSED - Dotter et al. 2008), with [$\alpha$/Fe]=0.4 and primordial helium.  The asymptotic giant branch (AGB) is also shown, based on BaSTI  isochrones (Pietrinferni et al. 2006), since they are not available in DSED.
The result is in very good agreement with results from Ortolani et al. (2006) within $1-\sigma$. We obtain a reddening of $E(B-V)=0.38\pm0.04$, a distance of d$_{\odot}=6.0\pm0.6$ kpc, and a metallicity of [Fe/H]$= -1.34\pm0.18$. Our age determination indicates an old age of $13.4^{+1.0}_{-1.2}$ Gyr, indicating that AL~3 is another relic fossil. It is important to stress that our distance of 6.0 kpc is also in agreement, within $1-\sigma$, with the value of 6.5 kpc from Harris (1996, 2010), Rossi et al. (2015), and Baumgardt et al. (2019). 

 Finally, we note that the dispersion of the data could be due to differential reddening, together with contamination and blends.
The reddening of AL~3 is relatively high, and the differential reddening is certainly an issue, as it is in other bulge globular clusters with similar reddening.
We expect an amount of about 20\% of differential reddening.
In principle it can be corrected, but the standard procedures for differential reddening correction in this cluster cannot be applied due to the few bona fide reference stars in the CMDs.

\begin{figure}
    \centering
    \includegraphics[width=\columnwidth]{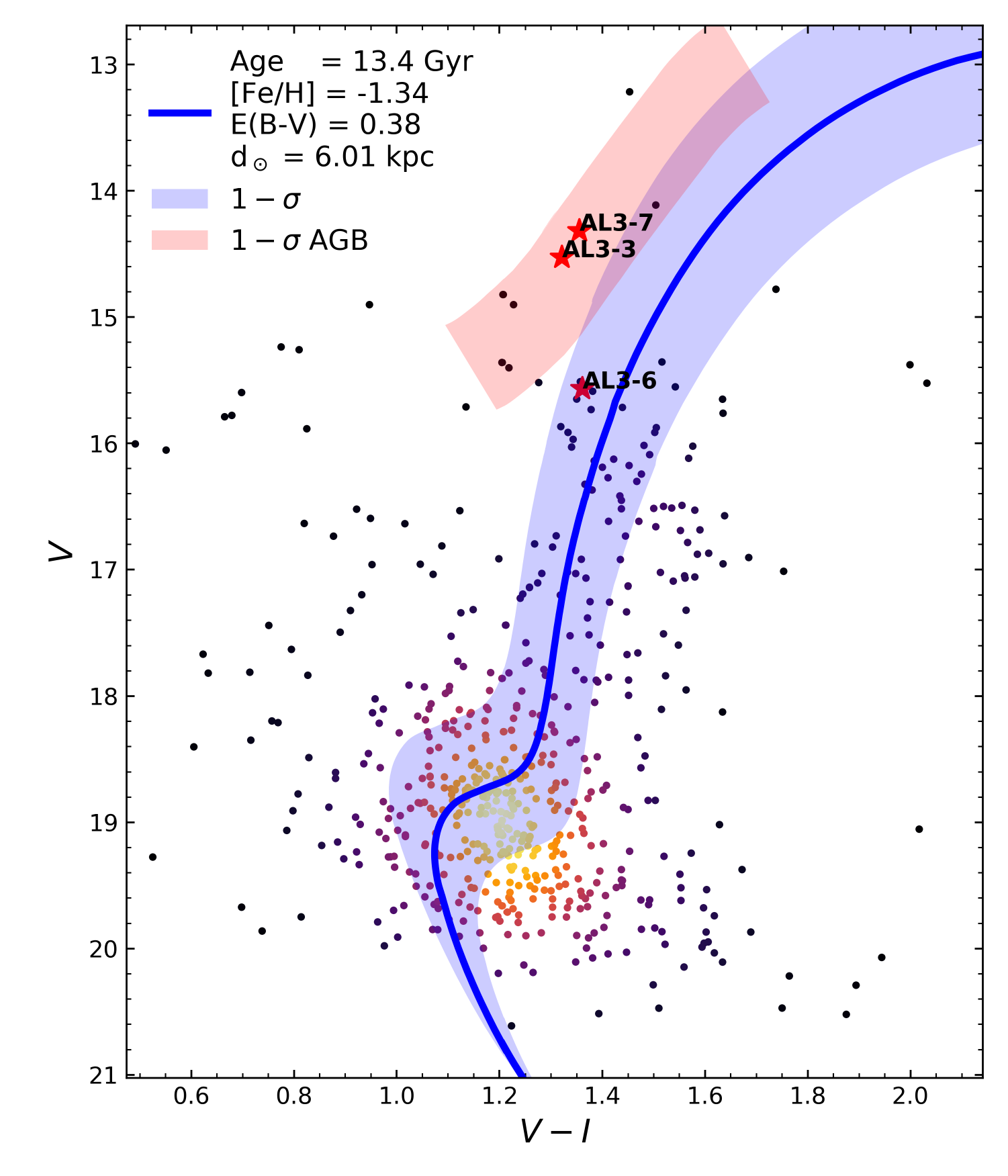}
    \caption{AL~3 $V$ vs. $V-I$ CMD. The black dots are the stars within 120 pixel from the cluster centre (see Ortolani et al. 2006). The red stars are the observed stars of the present work. The solid blue line represents the median solution of the isochrone fitting, while the blue region reveals the solutions within $1-\sigma$.}
    \label{fig:al3-cmd}
\end{figure}

Figure \ref{fig:al3-cmd} shows the solution of isochrone fitting. The solid blue line represents the median solution, while the shaded regions indicate the solutions within $1-\sigma$. The red stars are the three sample stars. Finally, Figure \ref{fig:al3-corner} exhibits the corner plots showing the (anti)correlations between the parameters.

\section{Stellar parameters}

\subsection{NGC 6558 and HP~1}

Individual stars of NGC~6558 were analysed
with high-resolution spectroscopy by 
Barbuy et al. (2007, 2018b) and with moderate-resolution spectroscopy
by Dias et al. (2015). The stars NGC6558-42 and NGC6558-64
are studied here.

Similar studies of HP~1 were carried out in Barbuy et al. (2006, 2016)
and Dias et al. (2016).
In the 2006 article, the bright red giants were labelled with numbers 1 to 6,
for the purpose of identifying them in the cluster chart.
In 2016, we adopted the identification numbers corresponding to the 
photometric reductions
relative to observations obtained at the New Technology Telescope (NTT)
at ESO, in 1994, as described in Ortolani et al. (1997).
HP1-4 and HP1-5 are stars 2115 and 2939 in Barbuy et al. (2016).
HP1-2 is the same as in Barbuy et al. (2006).
In our study, we only analysed HP1-5.

\subsection{AL~3}

The magnitudes and colours as follows are indicated in Table 3:
B, V from Ortolani et al. (2006), V, I from Rossi et al. (2015),
JHK from the 2MASS catalogue (Skrutskie et al. 2006),\footnote{
  $\mathtt{http://ipac.caltech.edu/2mass/releases/allsky/}$;
$\mathtt{https://irsa.iapc.caltech.edu}$}
and JHK from the VVV survey 
(Saito et al. 2012).\footnote{$\mathtt{horus.roe.ac.uk/vsa}$}

Effective temperatures were initially derived from  $B-V$, $V-I$, $V-K$, and $J-K$ using
the colour-temperature calibrations of Alonso et al.  (1999).
V,I Cousins were transformed to V,I Johnson using
 ($V-I$)$_{C}$=0.778($V-I$)$_{J}$    (Bessell 1979).
The $J,H,K_S$ magnitudes  and colours were  transformed from the 2MASS
system to California Institute of  Technology (CIT), and from this to
Telescopio Carlos S\'anchez (TCS),  using the relations established by
Carpenter (2001)  and  Alonso et  al.  (1998).
The conversion of JHK VVV colours to the JHK 2MASS system was done using
relations by Soto et al. (2013). 

\begin{figure*}[h!]
\centering
\includegraphics[angle=0,scale=.95]{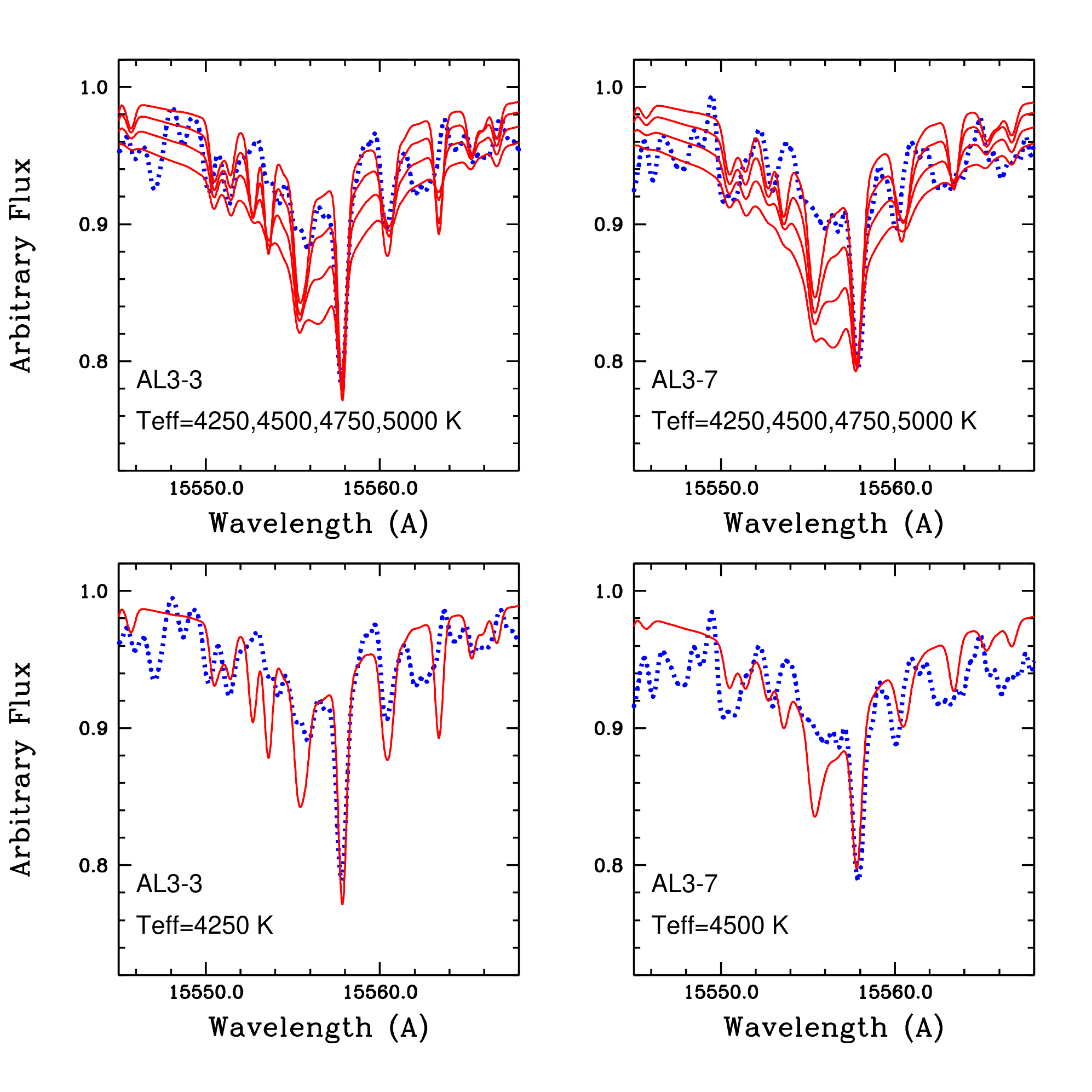}
\caption{AL~3-3 and AL~3-7: Hydrogen Brackett 16 line computed
  for T$_{eff}$ = 4250, 4500, 4750, 5000 K (upper panels) and adopted values of 4250 K  and 4750 K,
  respectively (lower panels).
The dashed line is the observed spectrum}
\label{hline1}
\end{figure*}

The temperatures resulting from photometry are of the order
of 5000 K for the three stars.
These temperatures, however, are not compatible with
another indicator, which is the Hydrogen Brackett 16
line, centred at 15556.457 {\rm \AA}. A fit of
this line for both AL~3 stars was carried out iteratively, after deriving their CNO abundances.
The resulting temperatures, adopted in the following analysis, are 4250 K and 4500 K for AL3-3 and AL3-7, respectively. The fits to the
hydrogen line are shown in Figure \ref{hline1}.
 For AL3-6, the low quality of the spectrum does not allow the fit of
the hydrogen line, in particular due to strong telluric
absorptions in the region. It appears to be cooler and compatible with 4150 K. This incompatibly between photometric and hydrogen-wing-derived temperatures is
a main source of uncertainty in the present study.

To derive the gravity, we used the PARSEC isochrones 
(Bressan et al. 2012).\footnote{http://stev.oapd.inaf.it/cgi-bin/cmd} To inspect the isochrones,
we adopted a metallicity of [Fe/H]$=-1.0$, or
overall metallicity Z$ = 0.00152$ (10 times below solar), and an age of 12 Gyr.
Assuming a reddening of $E(B-V) = 0.36$ (Ortolani et al. 2006, and present results), leading to $E(V-I) = 0.478$
and A$_V = 1.12$, we transformed the apparent magnitudes to absolute magnitudes,
as well as the colours ($V-I_{corr}$ = $V-I$-$E(V-I)$),
and we identified the correspondence of the observed stars to the theoretical isochrone.

 The metallicity resulting from the CMD fitting is [Fe/H]=-1.34, which was imposed as a prior. We inspected individual lines of Fe in the AL3-3 spectrum and the fits are more compatible with
[Fe/H]=-1.0. There is also the evidence from other similar bulge globular clusters such as NGC 6558, NGC 6522, HP~1, and Terzan 9, which are found to have
[Fe/H]$\sim$-1.0 from high-resolution spectroscopy. Bica (2016) showed that there is a peak in metallicity at
[Fe/H]$\sim$-1.0 in the bulge, which we also adopted for AL~3. An isochrone fitting with this higher metallicity was tried, but appeared difficult to converge.
This is a second source of uncertainty of the present study.
Final adopted stellar parameters for program stars,
and of the reference stars Arcturus (Mel\'endez et al. 2003) and
$\mu$ Leo (Lecureur et al. 2007), are reported in Table \ref{param}.

\begin{table*}
\small
\caption[3]{\label{logmag}
AL~3: coordinates, magnitudes, and colours of sample stars.}
\begin{flushleft}
\begin{tabular}{lrrrrrrrrrrrr}
\noalign{\smallskip}
\hline
\noalign{\smallskip}
 \hbox{Star}& \hbox{2MASS}&   \hbox{$\alpha$(J2000)} & \hbox{$\delta$(J2000)}&
\hbox{V}& \hbox{I} & \hbox{J} & \hbox{H} & \hbox{K} & \hbox{J} & \hbox{H} & \hbox{K} &\\
& & & &  &  &  &  \multispan2 {2MASS} & & \multispan2 {VVV} \\
\noalign{\vskip 0.2cm}
\noalign{\hrule\vskip 0.2cm}
\noalign{\vskip 0.2cm}
B11 & 18101902-3144506 & 18 10 19.01 & -31 44 50.64 & 15.902 & 14.275 & 13.136& 12.440 & 12.280 & 13.017 & 12.397 & 12.248 & \\
B64 & 18101803-3145435 & 18 10 18.03 & -31 45 43.55 & 15.703 & 14.180 & 13.064 & 12.456 & 12.277 & 13.055 & 12.529& 12.384 &   \\
B73 & 18102150-3145268 & 18 10 21.50 & -31 45 26.77 & 15.709 & 14.187 & 13.128 & 12.449 & 12.316 & 13.047 & 13.047& 12.313 & \\
F42 & --- & 18 10 17.65 & -31 45 38.93 & 16.054 & 14.442 & ---& --- & --- & --- & --- & --- & \\
F97 & 18101520-3146014 & 18 10 15.21 & -31 46 00.67 & 16.037 & 14.467 & 13.183 & 12.481 & 12.338 & --- & 12.503& 12.378 &      \\
\hline
HP1-2 & 17310585-2958354 & 17 31 05.60 & -29 58 34.00 & 16.982 & 14.332 & 12.210 & 11.268 & 10.969 &  14.588 & 13.675& 13.368&    \\
HP1-4 & 17310538-2959199 & 17 31 05.30 & -29 59 20.00 & 17.070 & 14.281 & --- & 11.67 & --- & 11.258 & 11.392& 10.688 & \\
HP1-5 & 17310729-2959021 & 17 31 07.20 & -29 59 02.00 & 17.131 & 14.395 & 11.901 & 10.869 & 10.595 & 12.021 & 11.285& 10.898 & \\
\hline
AL3-3 & 18140691-2839087& 18 14 06.90  & -28 38 09.0 & 14.524 &13.204 & 12.214 & 11.631 &11.469 & 12.211 & 11.714& 11.544 & \\
AL3-6 & 18140592-2838049& 18 14 05.80  & -28 38 06.0 & 15.563 &14.203 & 12.763 & 12.272  & 12.256 & --- & --- & --- & \\
AL3-7 & 18140529-2838168& 18 14 05.30 & -28 38 19.0 & 14.313 &12.963& 
11.878 & 11.296 & 11.170 & 11.920 &11.472 &     11.195 \\       
\noalign{\smallskip} \hline \end{tabular}
\end{flushleft}
\end{table*}

\begin{table}
\caption[4]{\label{param}
Adopted stellar parameters for individual stars in  NGC~6558,
AL~3, and HP~1, and resulting C, N, O abundances. For  NGC~6558, the stellar parameters are from Barbuy et al. (2007), and for HP~1 they are from Barbuy et al. (2016). Stellar parameters for the Sun, Arcturus, and $\mu$ Leo
are also included. }
\begin{tabular}{l@{}r@{}r@{}rrr@{}r@{}r@{}r@{}}
\noalign{\smallskip}
\hline
\noalign{\smallskip}
\hbox{Name}&
\hbox{T$_{\rm eff}$} &\hbox{\phantom{-}\phantom{-}log~g}& \hbox{\phantom{-}\phantom{-}[Fe/H]} &  \hbox{v$_t$} & [C/Fe] & [N/Fe] & [O/Fe] & \\
 & \hbox{(K)} & &  &  \hbox{km/s} & & & & \\ 
 \noalign{\smallskip}
\noalign{\hrule}
\noalign{\smallskip}
& \multicolumn{7}{c}{ Program stars } \\
\hline
F42 & \phantom{-}3800  & \phantom{-}0.5   & \phantom{-}$-$1.05 & 1.65 & $-$0.5 & +0.8 & +0.5&\\
HP1-5 & \phantom{-}4525  & \phantom{-}2.0  & \phantom{-}$-$1.07 & 1.55 & $-$0.1 & 0.8  &0.5 & \\
AL3-3 & \phantom{-}4250  & \phantom{-}1.7   &\phantom{-}$-$1.00  &1.2  & $-$0.1 & 0.0  &0.0 & \\
AL3-6       & 4150  & 1.3   &-1.00  &1.2   & +0.8 & +1.0  &+0.8 & \\
AL3-7  & \phantom{-}4500  & \phantom{-}1.5   &\phantom{-}$-$1.00  &1.2  & $-$0.2 & -0.1 &0.0 &  \\
\hline
\noalign{\smallskip}
& \multicolumn{7}{c}{ Reference stars } \\
\hline
Sun      & \phantom{-}5770 & \phantom{-}4.44 &  \phantom{-}+0.00  & 1.0  & +0.00  & +0.00 & +0.00 & \\
Arcturus & \phantom{-}4275 & \phantom{-}1.55 & \phantom{-}$-$0.54 & 1.65 & +0.11 & +0.40 & +0.43 & \\
$\mu$ Leo & \phantom{-}4540 & \phantom{-}2.3 & \phantom{-}+0.30 & 1.3 & $-$0.1 & +0.65 & +0.00 & \\
\noalign{\smallskip} \hline 
\end{tabular}
\end{table}

\begin{figure}[h!]
\centering
\includegraphics[angle=0,scale=.45]{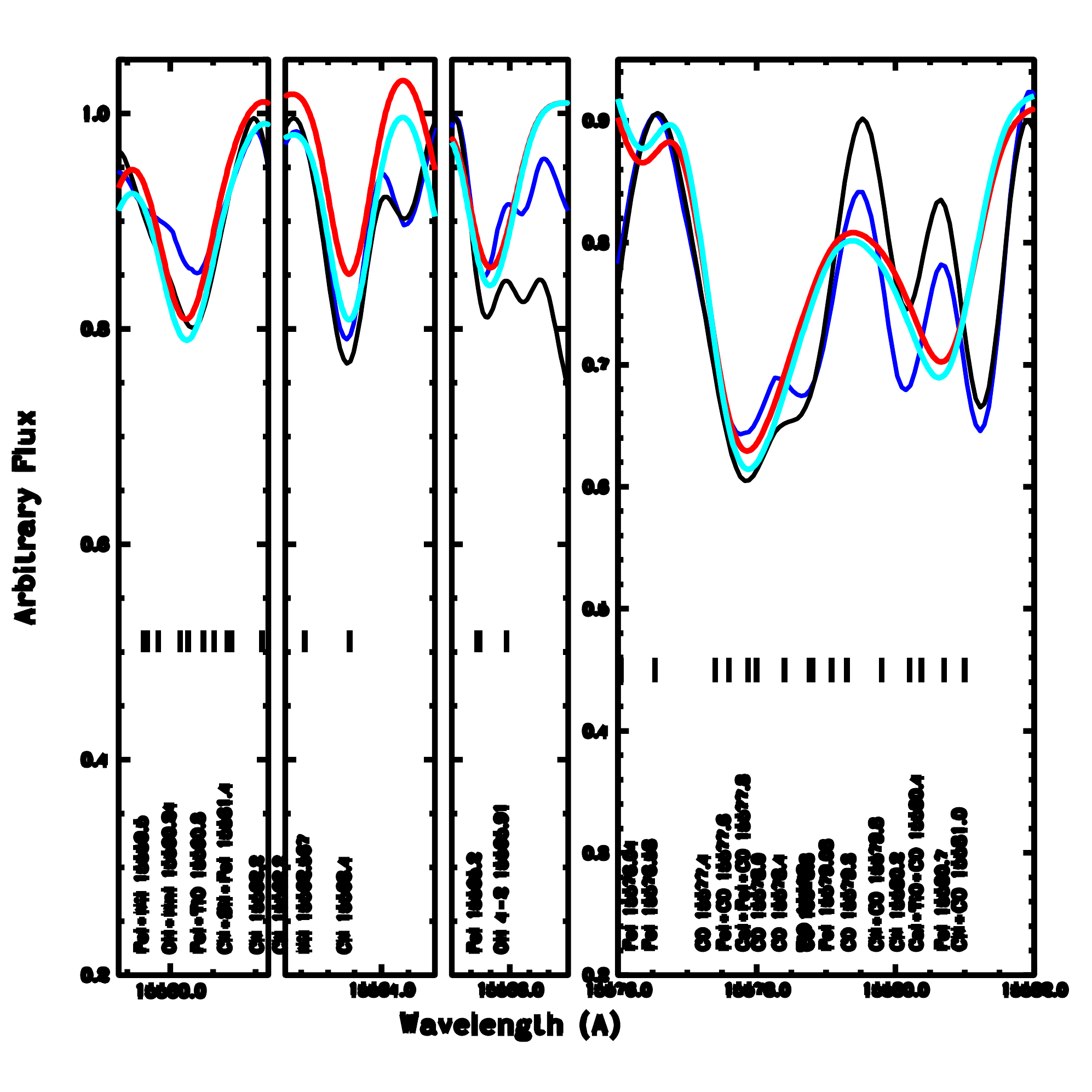}
\caption{AL~3-6: Tentative fit to this low S/N spectrum,
showing that it is compatible with a high CNO abundance.
Blue and black lines: Same observed spectrum normalized in 
two different ways. 
Red and cyan lines: Synthetic spectra computed
with [C/Fe]=0.7 (red), 0.8 (cyan), [N/Fe]=+1.0, [O/Fe]=+0.8.}
\label{al36cd}
\end{figure}

\section{CNO abundances}

The atmospheric models were interpolated in the grid of models by Gustafsson et al. (2008). 
The synthetic spectra were computed employing the PFANT code
described in Barbuy et al. (2018b).
In order to derive the C, N, O abundances, we 
fitted the CN, OH, and CO lines iteratively.

\subsection{The cool red giant N6558-42}

\begin{figure*}[h!]
\centering
\includegraphics[angle=0,scale=.90]{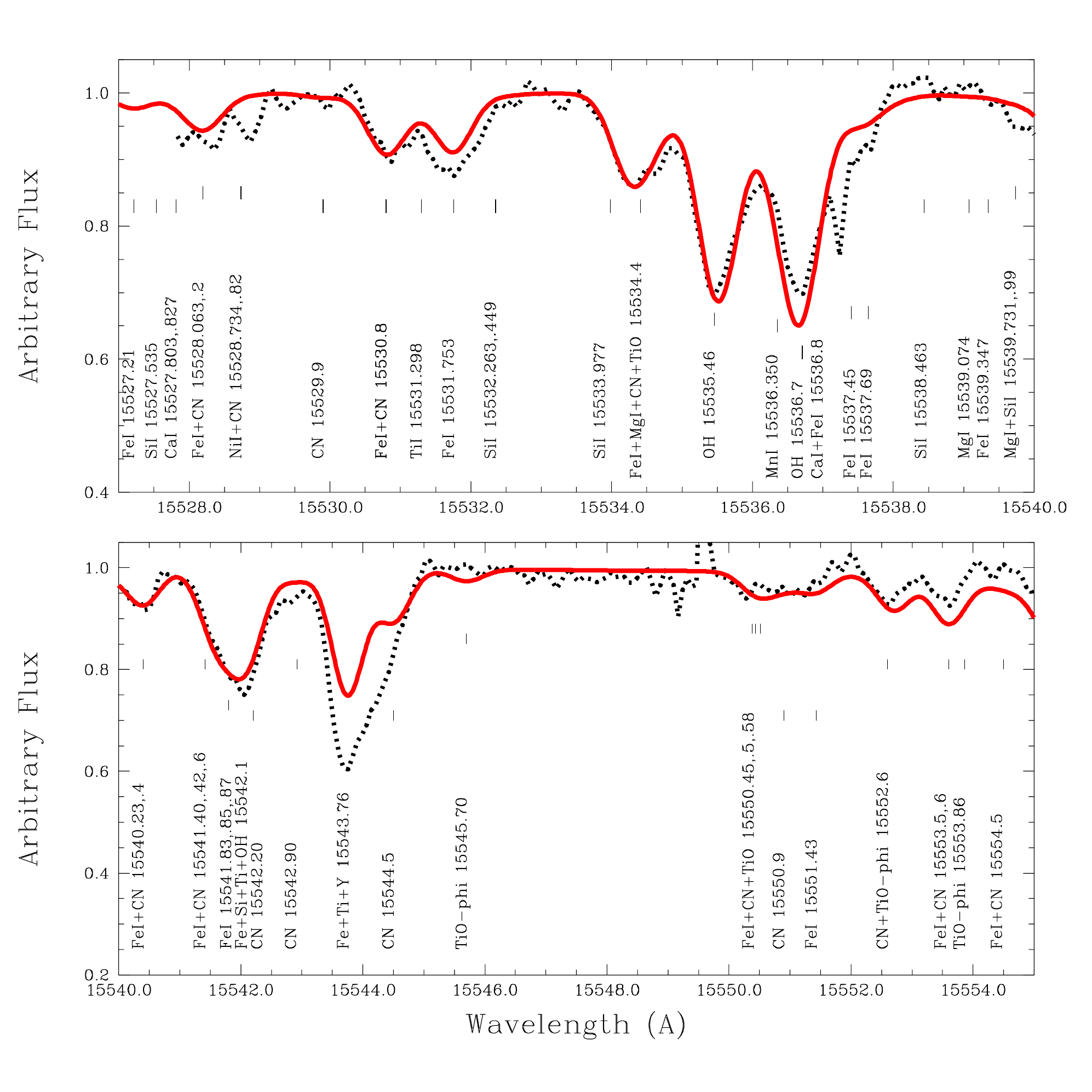}
\caption{NGC 6558-42: Line identification
in the range 15527-15555 {\rm \AA}. Dashed line: Observed spectrum. Solid red line: Synthetic spectrum.
Synthetic spectrum computed with [C/Fe]=-0.5, [N/Fe]=0.8, [O/Fe]=+0.5.}
\label{n655842ab}
\end{figure*}

\begin{figure*}[h!]
\centering
\includegraphics[angle=0,scale=.90]{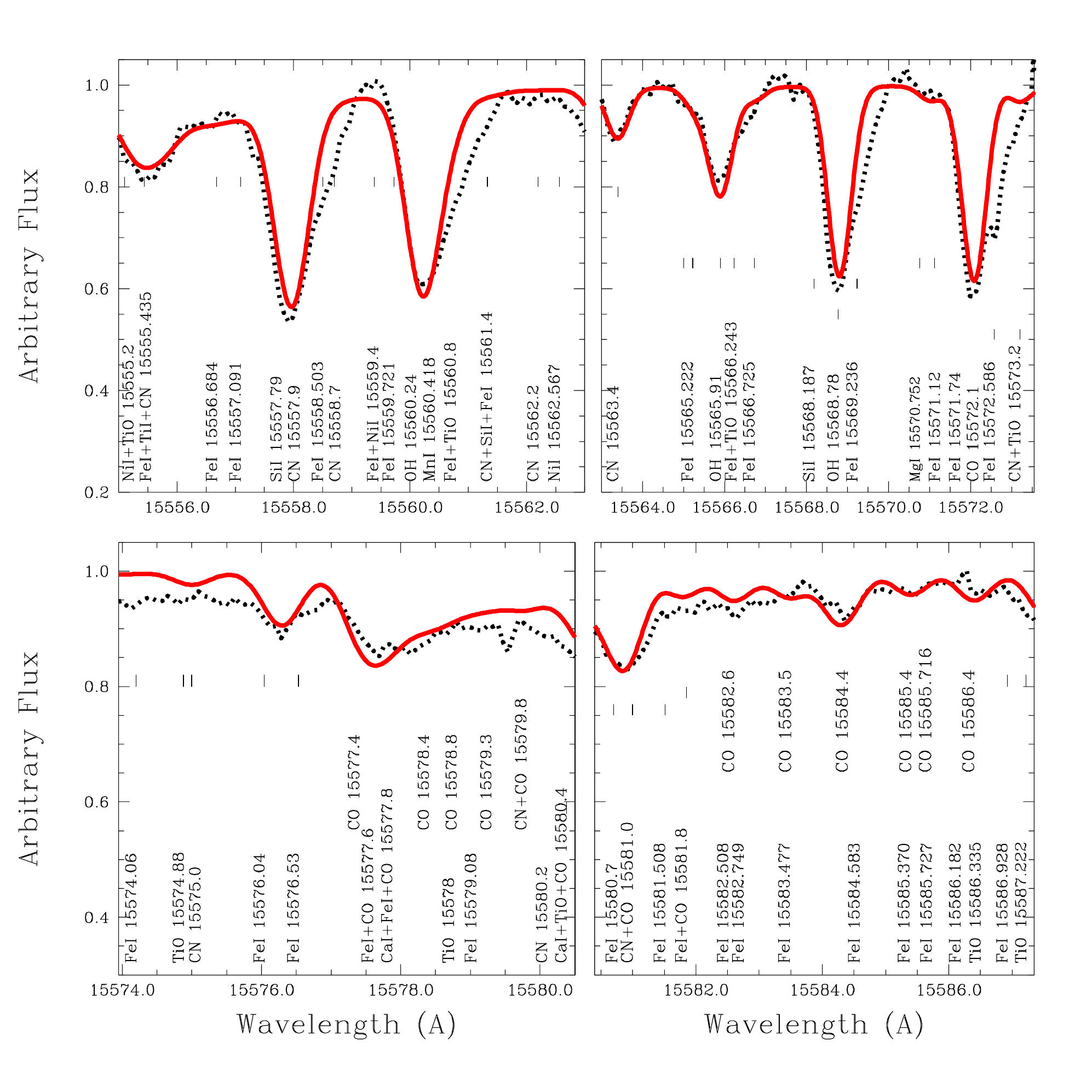}
\caption{NGC 6558-42: Same as Fig. \ref{n655842ab}, in the range 15555-15587 {\rm \AA}.}
\label{n655842cd}
\end{figure*}

The cool red giant, NGC6558-42, shows strong lines
and is a typical red giant. For this reason, we show the fits to the spectrum of this star in detail in Figures
\ref{n655842ab} and \ref{n655842cd}.

The star NGC6558-64 instead, which would have an effective
temperature of 4850 K according to the analysis from optical
spectra by Barbuy et al. (2018b), could be as hot as 5500 K.
This is seen from the profile of its Hydrogen Brackett 16 line;
however, this should be taken with caution due to defects in the observed spectrum.
For this reason, we could not converge on CNO abundances for this
star.

\subsection{AL-3 and HP~1}

\begin{figure*}
\includegraphics[angle=0,scale=.95]{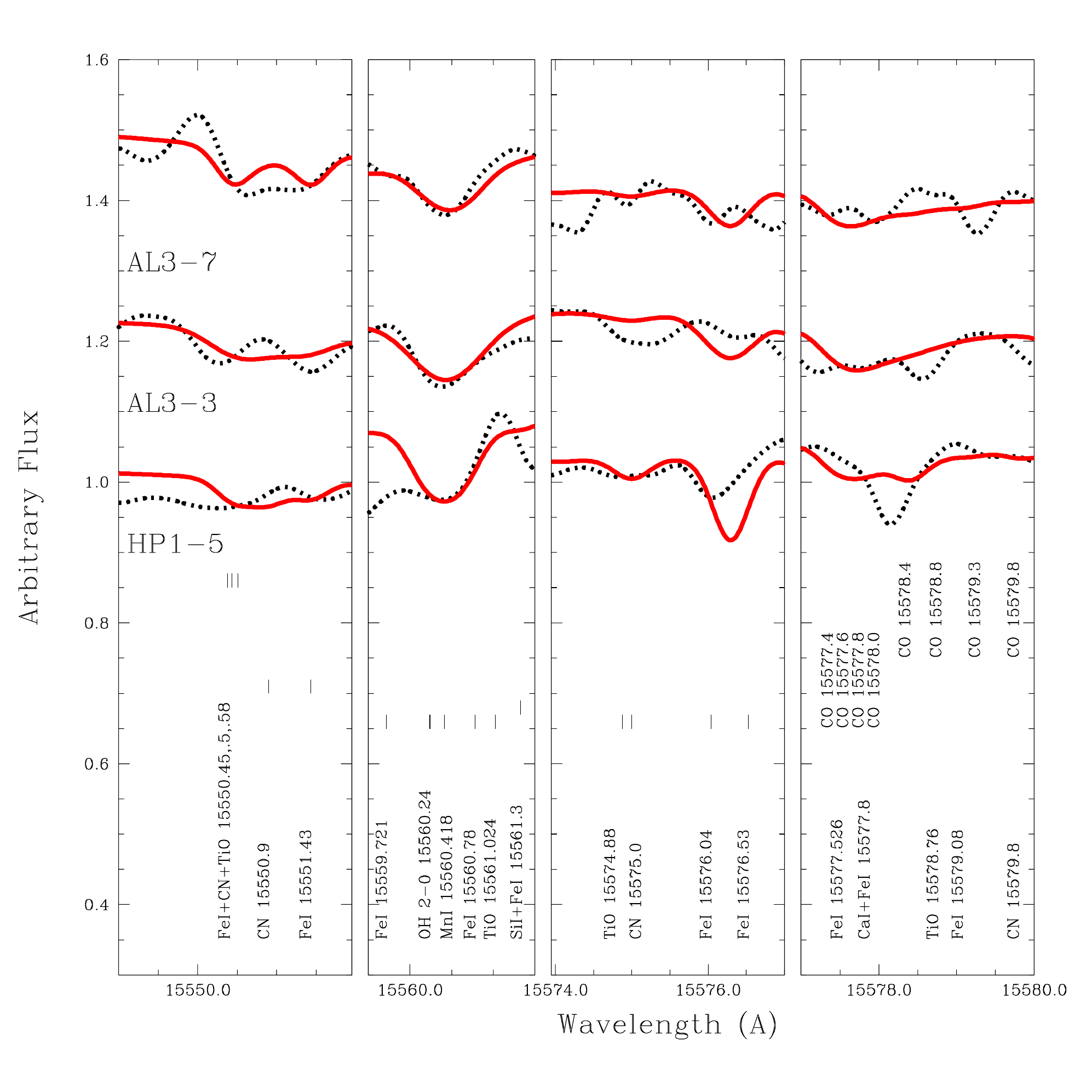}
\caption{HP~1-5, AL~3-3, and AL~3-7: Spectrum in selected wavelength regions containing CN, OH lines and the CO bandhead. 
Synthetic spectra are computed for the CNO abundances
given in Table \ref{param}.}
\label{hp15eal}
\end{figure*}

There is a clear contrast between the spectra of
HP1-5, AL3-3, and AL3-7, that have shallow lines, and AL3-6 and 
NGC6558-42, that show strong lines.
Whereas NGC6558-42 is a typical red 
giant, the stars AL3-3, AL3-7, and HP1-5 show weak molecular lines.
From the location of stars AL3-3 and AL3-7 in the CMD of Fig. \ref{fig:al3-cmd}, they should be AGB stars.
In Fig. \ref{hp15eal}, we show the observed spectrum in the selected
wavelength regions containing CN, OH lines, and the CO bandhead at 15578 {\rm \AA} for stars AL3-3, AL3-7, and HP1-5.
The molecular lines are very shallow, due to a combination of warm temperatures and
low metallicities. Clearly, the CNO abundances derived for these stars are less
reliable than for the cool star NGC6558-F42. 
Their CNO abundances are compatible with being close to solar,
but given the shallowness of the lines, it is clear that 
the molecular lines are not reliable for abundance measurements.

AL3-6 instead shows very strong CNO lines. Figure \ref{al36cd} indicates
that [C/Fe]= +0.7, 0.8, [N/Fe]=+1.0, [O/Fe]=+0.8 for this star.
We show two different renormalisations to illustrate the difficulty
in analysing this spectrum. Additionally, the computations with two different
carbon abundances illustrate the extreme sensitivity of the lines.
 Clearly, however, there is an urgent need to observe this star in the optical
and/or in a more extended wavelength region in the H-band to obtain firm conclusions on the CNO abundances of AL~3.

\subsection{Errors}

The main uncertainty in the derivation of CNO abundances in AL~3 stems from
the effective temperatures. Adopting a colour excess E(B-V)=0.36
the photometric magnitudes results in temperatures of $\sim$5000 K.
In order to be compatible with the wings of hydrogen lines,
we would have to adopt E(B-V)=0.2, but
the present fit of the CMD confirms the high reddening value.

Hydrogen wings that we adopted are a very good temperature indicator, and the temperatures can be roughly derived for the two
stars showing the line, whereas for the cooler star AL3-6, the H line is not strong, indicating that it is cooler than the other two stars. For this star, we adopted T$_{\rm eff}$ = 4150 K, compatible with its almost absent hydrogen-profile and a temperature compatible with its location in the RGB. With this temperature, we obtained high CNO values and could not converge with lower values.
We note that the CN, CO, and OH dissociation equilibrium interplay
gives a strong constraint on the result, but at the same time, these are very sensitive to abundance variations.
We adopted effective temperature errors of $\pm$250 K for AL3-6, and $\pm$150 K for 
AL3-3 and AL3-7, errors of $\pm$0.8 for the gravity,
and $\pm$0.3 for metallicity.

\subsection{Orbits of AL~3}

In order to investigate whether AL~3 belongs to the bulge, thick disc, or halo, given its solar CNO abundances, we carried out the calculations of orbits for the cluster. We employed a Galactic model that includes an exponential disc made by the superposition of three Miyamoto-Nagai potentials (Smith et al. 2015), a dark matter halo with a Navarro-Frenk-White (NFW) density profile (Navarro, Frenk \& White 1997), and a triaxial Ferrers bar. The total mass of the bar is $1.2 \times 10^{10}$ M$_{\odot}$, an angle of $25^{\circ}$ with the Sun-major axis of the bar, and a major axis extension of 3.5 kpc. We assumed three pattern speeds of the bar $\Omega_b= 40$, 45, and 50 km s$^{-1}$ kpc$^{-1}$. We kept the same bar extension, even though we changed the bar pattern speed. Our Galactic model has a circular velocity $V_0=241$ km s$^{-1}$ at $R_0=8.2$ kpc (Bland-Hawthorn \& Gerhard 2016; which is also compatible with the Gravity Collaboration et al. (2019)), and we assumed a peculiar motion of the Sun with respect to the local circular orbit of $(U,V,W)_{\odot}= (11.1, 12.24, 7.25)$ km s$^{-1}$ (Sch{\"o}nrich, Binney \& Dehnen 2010). 

To take into account the effects of the observational uncertainties associated with the cluster's parameters, we generated a set of of 1000 initial conditions using a Monte Carlo method considering the errors of distance, heliocentric radial velocity, and absolute proper motion components. With such initial conditions, we integrated the orbits forward for 10 Gyr using the NIGO tool (Rossi 2015).

\begin{figure*}
\includegraphics[angle=0,scale=.65]{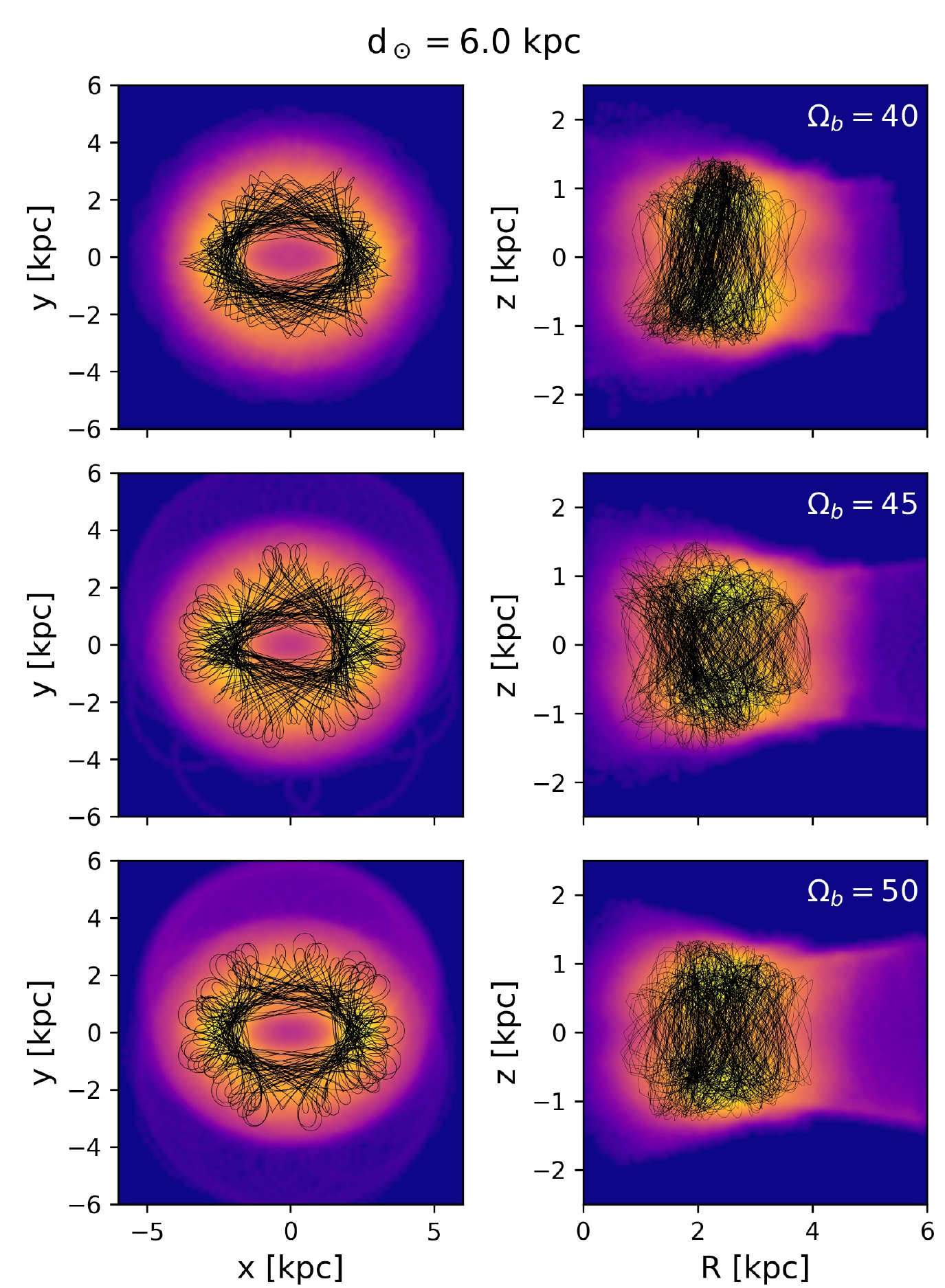}
\includegraphics[angle=0,scale=.65]{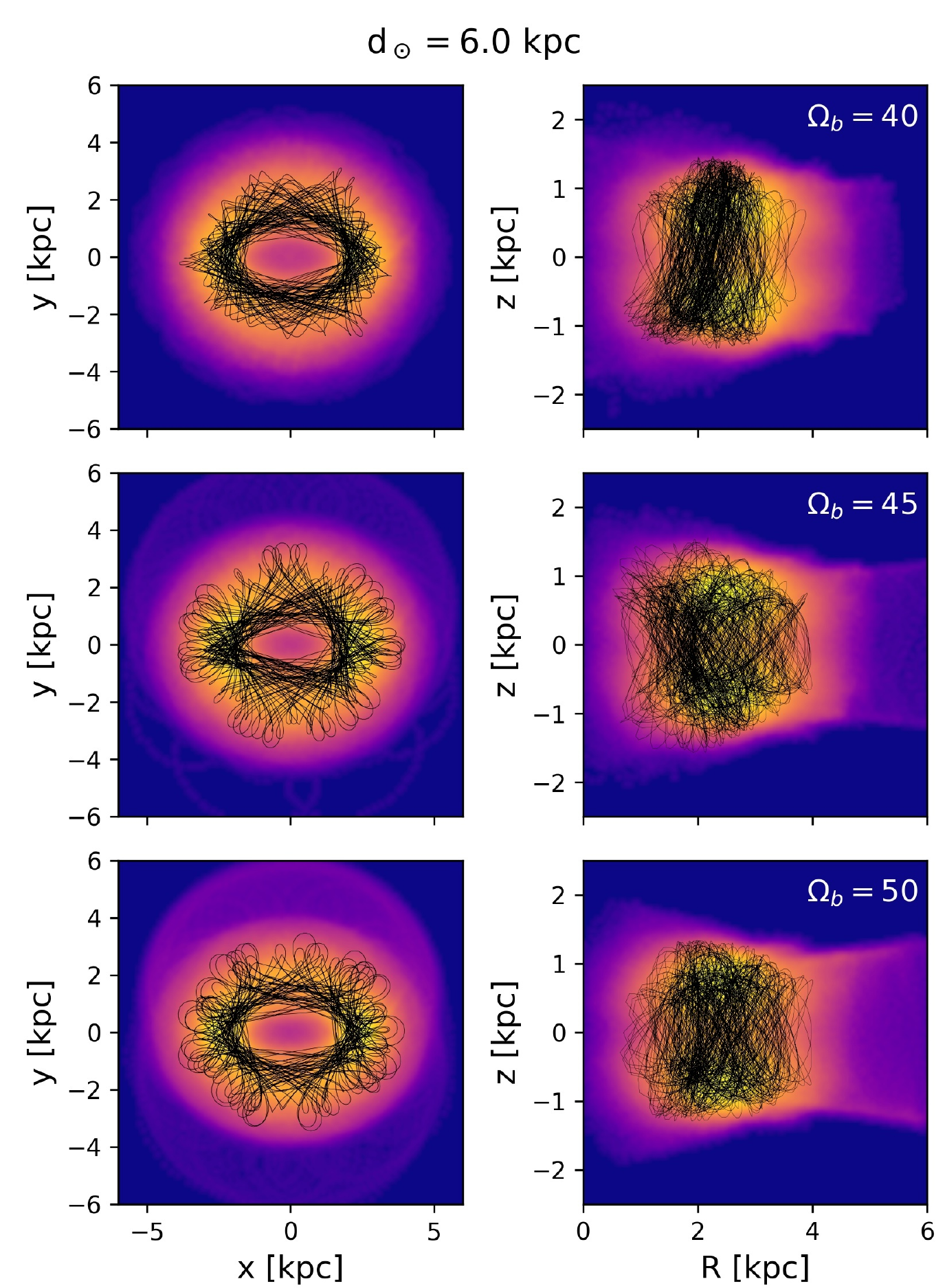}
\caption{Probability density map for $x-y$ and $R-z$ projections of the set of orbits for AL 3 for distances
of 6.0 kpc (left panels) and 6.5 kpc (right panels), using three different values of $\Omega_b= 40, 45,$ and $50$ km s$^{-1}$ kpc$^{-1}$. The orbits are co-rotating with the bar frame. Yellow corresponds to the larger probabilities. The black lines show the orbits using the central values.} 
\label{orbits}
\end{figure*}

\begin{figure*}
\includegraphics[angle=0,scale=.8]{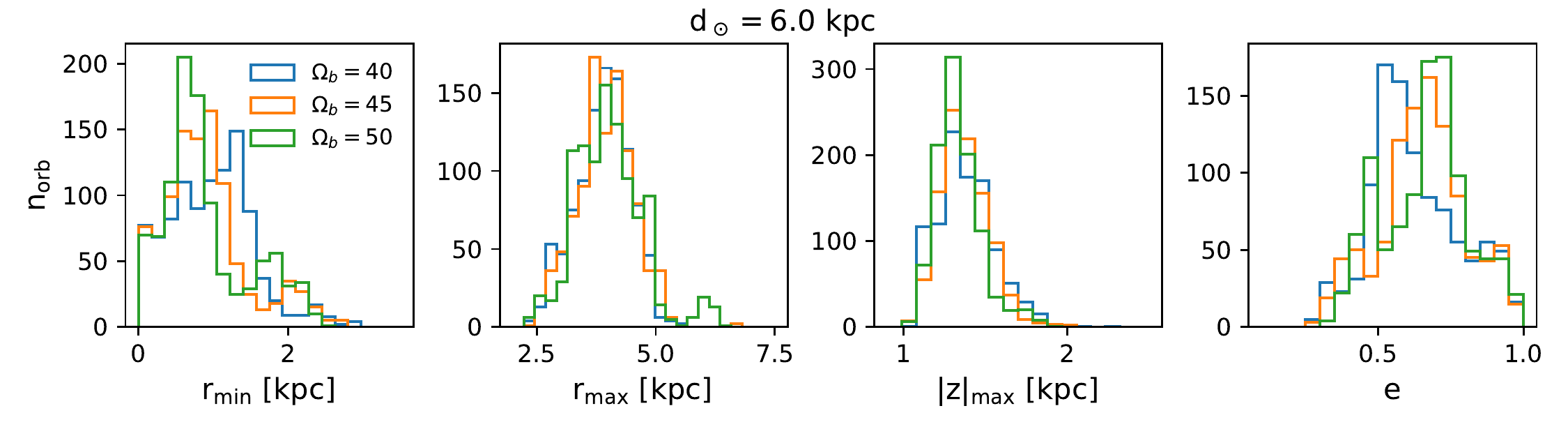}
\includegraphics[angle=0,scale=.8]{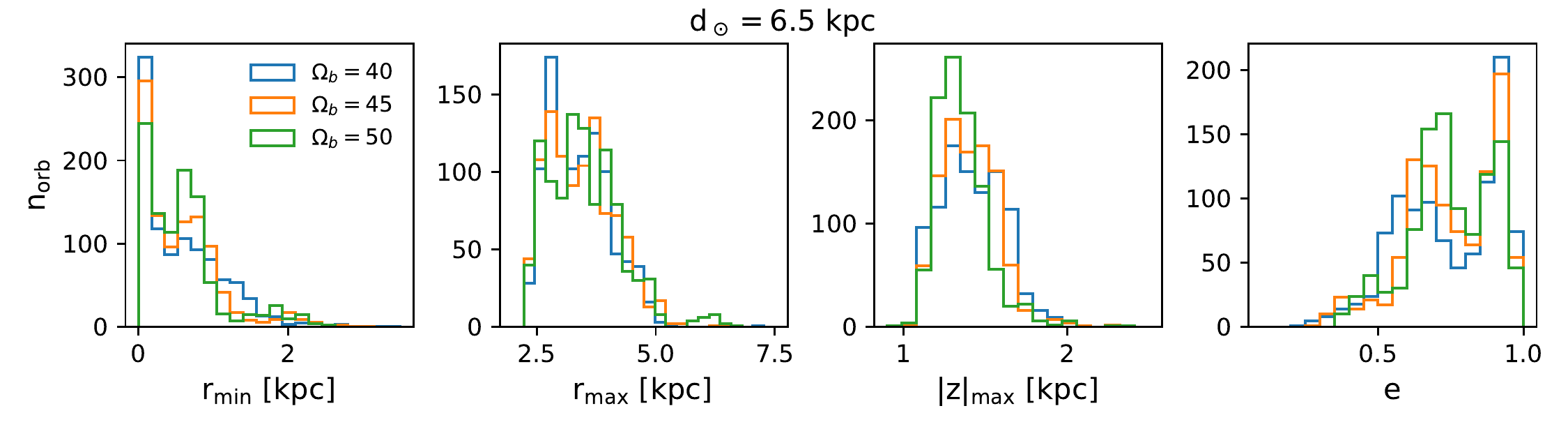}
\caption{Distribution of orbital parameters for AL 3, for distances of 6.0 kpc (top panels) and 6.5 kpc (bottom panels), with pericentric distance $r_{\rm min}$, apocentric distance $r_{\rm max}$, maximum vertical excursion from the 
Galactic plane $|z|_{\rm max}$, and eccentricity $e$. The colours show the different pattern speed of the bar, $\Omega_b=40$ (blue), 45 (orange), and 50 (green) km s$^{-1}$ kpc$^{-1}$.} 
\label{histograms}
\end{figure*}

In Fig. \ref{orbits}, we show the probability density map of the orbits of AL 3 in the $x-y$ and $R-z$ projections co-rotating with the bar, using the two adopted distances, 6.0 kpc (Ortolani et al. 2006) and 6.5 kpc (Baumgardt et al. 2019). The yellow colour displays the space region that the orbits of AL 3 cross more frequently. The black curves are the orbits using the central values of the observational parameters.
In Figure \ref{histograms}, we show histograms relating to the number
of probable orbits as a function of pericentric distance
(r$_{min}$), apocentric distance (r$_{max}$), maximum height above the plane (|Z|$_{max}$), and eccentricity ($e$). In Table \ref{Orb_para}, we give the median values of the orbital parameters, and the errors provided in each column are derived considering the 16th and 84th percentiles of the distribution. The orbital parameters are similar for the three pattern speeds.  
From the apocentric distance, we can see that AL~3 is mostly confined within $\sim 4$ kpc, and it has a high probability to belong to the bulge component ($\sim 88 \%$) when the adopted distance is 6.5 kpc. However, a significant fraction of orbits can reach apocentric distances up to $\sim6$ kpc, indicating that AL 3 could belong to the disc component, with a non-negligible probability ($\sim 12 \%$).

In Fig. \ref{histograms}, we show histograms relating
to the number of probable orbits as a function of
pericentric distance (r$_{min}$), apocentric distance
(r$_{max}$), maximum height above the plane
(|Z|$_{max}$), and eccentricity.
If a distance of 6.5 kpc is adopted, then
it is clear that AL~3 is a very central cluster,
with a maximum height reaching at most
$Z\sim1.5$ kpc, and with a high eccentricity orbit.
This is typical of the very old, moderately metal-poor
globular clusters of the Galactic bulge, similar
to NGC~6522, NGC~6558, and HP~1. 

On the other hand, if a distance of 6.0 kpc is adopted,
according to Ortolani et al. (2006), the cluster reaches distances farther from the Galactic centre with lower eccentricities (see left and bottom and top panels of Figures \ref{orbits} and \ref{histograms}, respectively).
The probabilities also change to an  $\sim$60 \% 
probability of belonging to the bulge and $\sim$40\%
to the disc. With this distance, the probability of being part of the disc increases significantly, and maybe this result could be more consistent with the solar CNO abundances.

\begin{table*}
\begin{center}
\caption{Median orbital parameters and membership probability of AL 3.}
\begin{tabular}{@{}ccccccc@{}} 
\hline
$\Omega_b$ & $\langle r_{\rm min}\rangle$ & $\langle r_{\rm max} \rangle$ & $\langle |z|_{\rm max}\rangle $ &$\langle e\rangle$ & $\mathcal{P}_{\rm bulge}$& $\mathcal{P}_{\rm disc}$\\
(km s$^{-1}$ kpc$^{-1}$) & (kpc) & (kpc) & (kpc) &  & (\%) & (\%)  \\
\hline
\multicolumn{7} {c} {$d_\odot= 6.0$ kpc} \\
\hline
40 & $1.00 ^{+0.43}_{-0.61} $ & $3.94 ^{+0.52}_{-0.66} $ & $ 1.36 ^{+0.19}  _{-0.15 }$ & $0.60^{+0.20}_{-0.10} $ & 61.7 & 38.3\\
45 & $ 0.84^{+ 0.47}_{-0.45} $ & $  3.97^{+0.55 } _{-0.65 }$ & $ 1.36 ^{+0.16}_{-0.13 } $ & $ 0.66 ^{+0.14 } _{-0.15} $ & 60.1 &39.9 \\
50 & $ 0.75^{+0.89 } _{- 0.34}$ & $ 3.98 ^{+0.72 } _{-0.65 }$ & $ 1.31 ^{+0.15 } _{-0.10 } $ & $ 0.68 ^{+ 0.12}_{- 0.20} $ & 60.6 & 39.4\\
\hline
\multicolumn{7} {c} {$d_\odot= 6.5$ kpc} \\
\hline
40 & $ 0.46^{+0.69 }_{-0.36 } $ & $ 3.34^{+0.69 }_{-0.61 } $ & $1.41  ^{+0.21 }  _{-0.19 }$ & $  0.75^{+0.18 }_{-0.19 } $ & 89.2 & 10.8\\
45 & $ 0.47^{+0.50 }_{-0.36 } $ & $ 3.39 ^{+0.76 } _{-0.67 }$ & $  1.40^{+0.17 }_{-0.17 } $ & $  0.76^{+0.17 } _{-0.15 } $ & 87.9 & 12.1\\
50 & $ 0.53^{+0.35 } _{-0.40 }$ & $ 3.42 ^{+0.75 } _{-0.73 }$ & $  1.33^{+0.15 } _{-0.17 } $ & $ 0.74 ^{+0.17 }_{-0.12 } $ & 88.5 & 11.5 \\
\hline
\label{Orb_para}
\end{tabular}
\end{center}
\end{table*}

\section{Conclusions}

We analysed spectra of individual stars of the globular clusters 
AL~3, NGC~6558, and HP~1, obtained with the PHOENIX spectrograph at the
Gemini South telescope. With a high spectral resolution of R$\sim$75,000,
in the H band centred at 15555 {\rm \AA}, the wavelength coverage is short (15520 - 15590 {\rm \AA}). 

In AL~3,
this limited wavelength range means that it is difficult to use atomic lines
to deduce the stellar parameters effective temperature, gravity, and metallicity.
For this reason, 
we obtained the effective temperature from the Hydrogen Brackett 16 line,
gravity from photometric data, and isochrones. The metallicity 
[Fe/H]$\sim$-1.3$\pm$0.3 was deduced from the observed CMD given in Ortolani et al. (2006) and is confirmed in the present work
through analysis of the same CMD.
 We note that we adopted [Fe/H]=-1.0 for the analysed stars,
due to spectroscopic evidence.
For NGC~6558 and HP~1, the stellar parameters were adopted
from previous analyses from optical spectra (Barbuy et al. 2007, 2018),
and Barbuy et al. (2006, 2016) respectively.
Adopting these stellar parameters, we computed the synthetic spectra in order
to derive the abundances of C, N, and O. Since they vary interdependently,
the fit was done iteratively, where particular attention was given to the
 CO bandhead.
 The stars analysed in NGC~6558 and HP~1 show typical CNO abundances of 
 red giants, and confirm previous oxygen abundance derivation.

AL~3 is a more complex case:
 two stars analysed in AL~3 show solar CNO abundance ratios, but based on
 very shallow lines, and the location of these two stars
 in the CMD point to them being AGB stars. The star AL3-6 shows instead very strong
 CNO abundances of the order of [C/Fe]=+0.8, [N/Fe]=+1.0, [O/Fe]=+0.8.
  A strong CNO abundance indicated by this cooler star shows that AL~3
 appears to be an extremely interesting old cluster.
 In conclusion, further investigation of this cluster are clearly needed.
 We also derived 
 the cluster's radial velocity, which in turn allowed us to compute the cluster orbits. For 
 the two AGB stars, we found a higher velocity, whereas
 for the third cooler star, the radial velocity is compatible with the value from
 Baumgardt et al. (2019).
 
 The orbital behaviour of AL~3 indicates that it is a
 typical inner bulge, moderately metal-poor globular cluster if its
 distance to the Sun is 6.5 kpc (Baumgardt 2019), but if the
 distance is 6.0 kpc (Ortolani et al. 2006 and present work), 
 there is an increased probability of AL~3 belonging to a diss component.
 We derive a very old age for AL~3 of 13.4 Gyr, which would point towards
 an old bulge cluster.
 Therefore, we conclude that the cluster AL~3 appears to
 be an extremely interesting cluster that should be
 further investigated through more wavelength-extended spectra,
 and including larger samples of member stars.
 
\begin{acknowledgements}
BB, HE, RR, JM, and EB acknowledge grants from FAPESP, CNPq and CAPES - Financial code 001.
TM acknowledges FAPESP postdoctoral fellowship no.
 2018/03480-7.
APV acknowledges FAPESP postdoctoral fellowship no.
2017/15893-1.
Support for MZ is provided by Fondecyt Regular 1191505, the BASAL CATA Center
for Astrophysics and Associated Technologies through grant PFB-06, and ANID -
Millennium Science Initiative Project ICN12\_009, awarded to the Millenium
Institute of Astrophysics (MAS).
S.O. acknowledges the partial support of the research
program DOR1901029, 2019, and the project BIRD191235, 2019 of the University of Padova.
\end{acknowledgements}


\begin{appendix}

\section{Isochrone fitting - corner plot}

\begin{figure}[h!]
    \centering
    \includegraphics[width=\columnwidth]{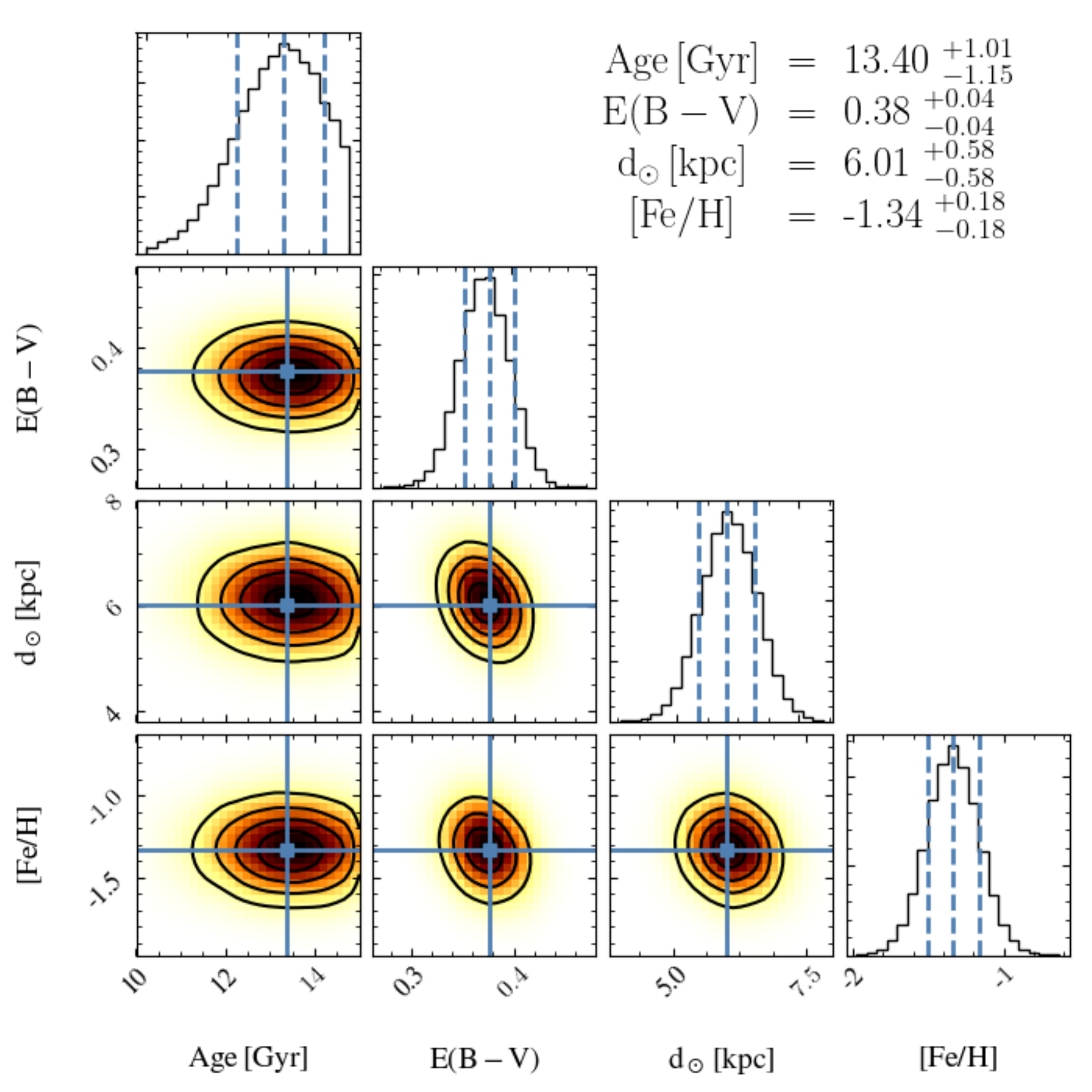}
    \caption{Corner plots representing the 4D parameter space of the
      Markov chain
      constructed through the Monte Carlo algorithm for the isochrone fitting.
      To represent the solutions, we adopt the peak of the distribution as
      the best value, and the uncertainties are computed using the
      $16^{th}$ and $84^{th}$ percentiles. More details are available in
      Souza et al. (2020).}
    \label{fig:al3-corner}
\end{figure}

The \texttt{SIRIUS} code performs the isochrone fitting through the construction of the Markov chain Monte Carlo (MCMC) algorithms. The purpose of the MCMC method is to obtain the posterior distribution of each parameter via the generation of random samples. Figure \ref{fig:al3-corner} shows the corner plots, that is, a graphical representation of the 4D parameter space of the isochrone fitting (age, reddening, distance, and metallicity). The histograms represent the cumulative best solutions (posterior distribution) for each parameter, while the 2D density maps show the correlations between the parameters as well as the region with the highest density. To represent the posterior distributions of each parameter, we adopt the region of highest density as the central value and the uncertainties calculated from the 16th and 84th percentiles.

\end{appendix}

\end{document}